\documentclass[runningheads]{llncs}
\usepackage[T1]{fontenc}
\usepackage{graphicx}
\usepackage{subcaption}
\usepackage{array}
\usepackage{amsmath}
\newcommand{\mb}{\textit{ReflectEd}}

\begin{document}
\title{\parbox{\textwidth}{\raggedright\small AIED 2026 Preprint}\vspace{0.4em}\\\large\mb: Evaluating Reflection-Driven Learning in an AI-Assisted System}
%
%
\author{Md Nazmus Sakib\inst{1}\orcidID{0009-0003-8282-3931} \and
Ishika Tarin\inst{1}\orcidID{0000-0003-0428-1447} \and
Naga Manogna Rayasam\inst{1} \and
Manas Gaur\inst{1}\orcidID{0000-0002-5411-2230} \and
Sanorita Dey\inst{1}\orcidID{0000-0003-3346-5886}} 
\authorrunning{M. N. Sakib et al.}
%
\small\institute{University of Maryland Baltimore County, Baltimore, MD 21250, USA
\email{\{msakib1,qp45654,tp34657,manas,sanorita\}@umbc.edu}}
\maketitle   
%
\begin{abstract}
In collaborative settings, sustaining momentum and engagement between checkpoints (e.g., meetings) can be challenging, often leading to task drift and reduced preparedness. To address this gap, we developed \mb, an AI-assisted system that supports between-checkpoint reflection through theory-driven prompts with progressively structured levels and mechanism-based scaffolding. We evaluated \mb~in a mixed-method study comparing two reflection configurations: a regular reflection workflow and a deeper reflection workflow that included an additional transformative reflection activity. Across conditions, participants reported steady engagement early in the week. In the deeper configuration, later reflections tended to exhibit higher actionability and richer forward-looking planning, while also being harder to sustain and more effortful during periods of active work. Partner-visible reflections were frequently described as supporting coordination by surfacing differences in focus and facilitating accountability. Overall, the findings characterize trade-offs between reflection depth, feasibility, and perceived preparedness for subsequent checkpoints. We discuss implications for the design of AI-assisted systems that support collaboration readiness and reflection-oriented regulation in time-constrained collaborative workflows.

\end{abstract}
\section{Introduction}
Academic project work often unfolds across multiple days, requiring students to sustain progress, respond to setbacks, and translate plans into follow-through. In these scenarios, learning is reflected not only in the final artifact, but also in how learners interpret experience, revise strategies, and adjust subsequent actions as part of ongoing self-regulation \cite{boud2013reflection}. Reflection supports this sensemaking process as prior work shows that guided reflection can promote understanding by prompting learners to revisit what happened and why \cite{zhang2025following,saucerman2017automating}. Learning is also a key outcome of reflection-on-action, where learners revisit completed work to inform future decisions \cite{munby1989reflection}. However, reflection can be difficult to sustain during active project work because it competes with the immediate goal of making progress \cite{xu2025productive}.
This challenge is especially visible between scheduled checkpoints (e.g., meetings). While meetings help learners align on goals and divide work, much of the learning-relevant progress occurs asynchronously, when individuals must maintain momentum and return prepared with meaningful updates. During this gap, attention shifts and task drift becomes more likely \cite{mark2015focused}. Temporal Motivation Theory suggests that motivation increases as deadlines approach and weakens when goals feel distant \cite{steel2006integrating}. This can lead students to postpone difficult work and arrive underprepared, particularly without lightweight support that helps them reflect and re-plan during this interval.\\
Prior research in HCI has explored meeting-centered supports such as summaries, dashboards, and reflective interventions to help groups revisit decisions and strengthen coordination \cite{samrose2021meetingcoach,wang2024meeting,park2023retrospector}. Recent work also highlights the value of AI-assisted goal reflection during meetings \cite{chen2025we,scott2024mental,scott2025does}. However, these systems largely focus on reflection during the meeting itself and offer limited support for reflection-on-action during the between-meeting interval, when learners must interpret progress, surface unfinished work, and decide next steps with little guidance \cite{wang2024meeting,park2023retrospector}. Reflection can also be difficult to sustain in real workflows because it adds effort on top of time-pressured tasks \cite{cho2022reflection,xu2025productive}. Reflection is most valuable when it supports learning that carries into subsequent action \cite{munby1989reflection}, yet it varies in form and depth \cite{fleck2010reflecting} and may be more effective when timed and scaffolded to fit ongoing work \cite{bentvelzen2022revisiting}. To better understand design tradeoffs in this space, we ask: \textbf{RQ: How do different levels of AI-scaffolded reflection between checkpoints affect learners' reflective engagement, collaboration experience, action planning, and preparedness for subsequent work?}\\   
To answer this question, we designed \mb, an AI-assisted system that supports structured reflection between two checkpoints or meetings in academic project workflows. After the first checkpoint, \mb~generates a meeting-grounded summary and uses it to deliver lightweight, personalized reflection prompts across the week. The design draws on reflection-on-action \cite{munby1989reflection} and prior HCI work showing that reflection varies in depth \cite{fleck2010reflecting}, and that reflective technologies benefit from concrete scaffolding \cite{bentvelzen2022revisiting,cho2022reflection}. By distributing prompts across days and grounding them in recent task context, \mb~aims to support sensemaking and action planning during the between-meeting interval while keeping added effort low \cite{xu2025productive}. This paper contributes (1) \mb, an AI-assisted reflection-based system for academic workflows between checkpoints, (2) an empirical comparison that characterizes tradeoffs between two reflection structures with different levels, and (3) design implications for AIED on supporting reflection-on-action in time-constrained collaborative learning.

\section{Related Works}

\subsection{Learning Through Reflection and Reflection-on-Action}
Reflection is often described as a meta-cognitive activity of examining one’s own thinking \cite{sandars2009use}. Schön’s Theory of Reflective Practice distinguishes between reflection-in-action and reflection-on-action. Reflection-in-action occurs during an activity through real-time thinking and adaptation, whereas reflection-on-action occurs after an activity and involves deliberate consideration of what happened and why to inform future practice \cite{Schon01071986}. Building on this literature, our work focuses on reflection-on-action in project-based learning contexts where learners revisit prior work, interpret incomplete progress, and plan subsequent actions. 

Reflection supports learning by helping learners interpret experience, monitor progress, and plan next steps. Prior studies show that reflective activities can support task understanding and encourage more deliberate information seeking \cite{zhang2025following,saucerman2017automating}. Researchers also highlight that awareness during activity can shape later reflection-on-action \cite{larsen2016using}, and that analyzing open-ended student reflections can reveal knowledge-building processes over time \cite{li2025reflections}. AI-based reflective agents have been shown to improve reflection quality across dimensions such as content specificity and depth \cite{pishtarireflectionapp}. Complementary work emphasizes transparency, user control, and recommendation strategies for supporting engagement, agency, and self-regulated learning in AI-supported educational systems \cite{barria2025using}. Together, this literature frames reflection as an intervention that can be designed to support sensemaking and regulation over time.

Many reflective learning designs draw on Experiential Learning Theory (ELT), which frames learning as a cycle of concrete experience, reflective observation, abstract conceptualization, and active experimentation \cite{kolb1984experiential}. In collaborative settings, reflection can also support coordination by helping group members align on progress, responsibilities, and next steps. Prior work on AI-supported collaborative learning examines systems that scaffold interaction quality, provide structured feedback, and support coordination \cite{soller2001supporting,is2007supporting}. Building on this direction, our work examines structured reflection as a lightweight scaffold that sustains coordination and self-regulation between checkpoints in short, time-bounded collaborative tasks.

\subsection{Intelligent Systems and Human-Centered AI in Collaborative Work}
Intelligent systems have long been explored to support collaborative learning and teamwork by improving interaction processes and providing targeted assistance \cite{mclaren2010supporting}. However, reflective learning support is difficult to operationalize in authentic workflows, particularly when reflection becomes effortful, repetitive, or disconnected from task demands. As a result, reflection-support systems must balance structure with feasibility by offering guidance that is useful, while remaining lightweight enough to fit ongoing work.

Prior research proposes concrete design perspectives for reflection-support technologies. Fleck and Fitzpatrick describe reflection as a spectrum of levels that vary in depth, noting that deeper reflection typically requires stronger scaffolding \cite{fleck2010reflecting}. Bentvelzen et al. articulate four design resources for reflective technologies: temporal perspective, discovery, comparison, and conversation \cite{bentvelzen2022revisiting}. These mechanisms highlight time-based framing, opportunities to notice change, and guided dialogue with peers or systems. Cho et al. identify gaps between reflection theory and interface practice, arguing that many systems label activities as reflective without providing sufficient structure to support sustained meaning-making \cite{cho2022reflection}.

Our work builds on these perspectives by operationalizing staged reflection levels and mechanism-based supports within a collaborative workflow. Rather than assuming reflection occurs naturally, we examine how AI-assisted reflection can provide temporally distributed scaffolding that helps learners maintain momentum, align with teammates, and prepare for subsequent checkpoints. In doing so, we position reflection as a designable component of collaborative learning systems that supports regulation and preparedness between moments of synchronous coordination.
\section{Methodology}

\paragraph{\textbf{Theoretical Baseline of \mb:}} Based on prior work on technology-supported reflection, we designed the reflection structure in \mb. Fleck et al. \cite{fleck2010reflecting} describe reflection as varying in depth and strength, while Bentvelzen et al. \cite{bentvelzen2022revisiting} outline mechanisms for designing reflective technologies through concrete system support. Cho et al. \cite{cho2022reflection} further argue that reflection is often treated as a design label rather than a scaffolded process, where systems assume reflection will occur without sufficient structure or opportunities for user-driven meaning-making. In practice, reflection can be difficult to sustain in authentic workflows because it adds cognitive effort on top of task demands, and repeated open-ended prompts often lead to shallow responses or disengagement. To reduce this burden and make reflection more learning-friendly, \mb~distributes structured activities across multiple days and gradually progresses from lightweight prompts to deeper, transformative reflection through level-based scaffolding and targeted guidance that supports ongoing sensemaking.

\textit{\textbf{Reflection Conditions:}} We employed two reflection conditions (RCs) and examined their impact across two participant groups. The RCs incorporated four reflection levels, ordered by depth: (1) Descriptive, (2) Explanatory, (3) Relational, and (4) Transformative \cite{fleck2010reflecting}. Transformative reflection is critical for learning because it supports deeper meaning-making and changes how learners act in future situations. We therefore designed a regular condition without the Transformative level and a deeper condition that included all four levels. In addition, both RCs drew on four reflection mechanisms: (1) Temporal, (2) Discovery, (3) Comparison, and (4) Conversation \cite{bentvelzen2022revisiting}. As shown in Table~\ref{tab:reflection_prompts}, reflective prompts were delivered on alternate days to reduce cognitive burden and to provide participants with time to make progress between reflections. The only exception was Transformative reflection, which was provided on Day~6 in the deeper condition.

\begin{table}[t]
\centering
\caption{Reflection workflow (Regular vs.\ Deeper). Days share the same prompt set, except Day~6 which is only included in the Deeper condition.}
\label{tab:reflection_prompts}
\scriptsize
\setlength{\tabcolsep}{5pt}
\renewcommand{\arraystretch}{1.2}

\begin{tabular}{p{0.14\columnwidth} p{0.32\columnwidth} p{0.50\columnwidth}}
\hline
\textbf{Day} & \textbf{Level \& Mechanism} & \textbf{Prompts \& Purpose} \\
\hline

Day 0 (CP1) 
& \textit{First Checkpoint}
& --- \\

Day 1
& \textbf{Descriptive}
& \textit{Comprehensive Summary:} Recap the meeting and briefly report progress so far. \\

Day 2
& ---
& --- \\

Day 3 
& \textbf{Explanatory} (Discovery + Conversation)
& \textit{Explain and Guide.} 
Discovery: (1) What was hard today and why? (2) What did you notice that you didn’t expect? \newline
Conversation (with \mb, up to 3 turns): one targeted question with up to two follow-ups to help turn challenges into guidance for next steps. \\

Day 4
& ---
& --- \\

Day 5 
& \textbf{Relational} (Comparison)
& \textit{Explore and connect.}
(1) Better/same/worse than yesterday? (2) Why? \newline
(3) Compared to your partner, what did you focus on most today? \textit{(Deeper RC only)} \newline
Supports self-comparison over time and partner comparison when available. \\

Day 6
& \textbf{Transformative} (Temporal)
& \textit{Take action and change.}
(1) What is one thing you will do differently next week? \newline
(2) What is one thing you will keep the same? \newline
(3) What is one task you will finish before the final checkpoint? \\

Day 7 (CP2) 
& \textit{Second Checkpoint}
& --- \\

\hline
\end{tabular}
\vspace{1pt}
\textit{\textbf{Progression labels:} Descriptive $\rightarrow$ Explanatory $\rightarrow$ Relational $\rightarrow$ Transformative. \textbf{Note:} Each checkpoint (CP) marks when users meet to share updates.}
\end{table}

\textit{\textbf{Study Design:}} Participants completed a collaborative poster design task over seven days to examine how structured reflection supports sustained collaboration across multiple days. The study was conducted remotely, and participants were instructed not to use any LLM tools during the task. Participants were allowed to form pairs based on preference; otherwise, they were randomly paired. Each pair held two online meetings: the first meeting focused on selecting a poster topic and dividing responsibilities, and the second meeting focused on reviewing and finalizing the poster. Each meeting was recommended to last no longer than 40 minutes. We provided example topics (e.g., reducing food waste, improving study habits, reducing screen time), while allowing pairs to choose their own. Between the two meetings, participants received notifications on alternate days through \mb~and completed reflection activities within the system. On reflection days, participants were asked to complete the activity within 24 hours of receiving the notification. In the deeper condition, participants could view their partner’s Day~3 reflection on Day~5 to support comparison and coordination, while \mb~conversations remained private. The study focused on how structured reflection influenced collaboration and preparedness, rather than evaluating the quality of the final poster.

\begin{figure}[h!]
    \centering
    \includegraphics[width=1\linewidth]{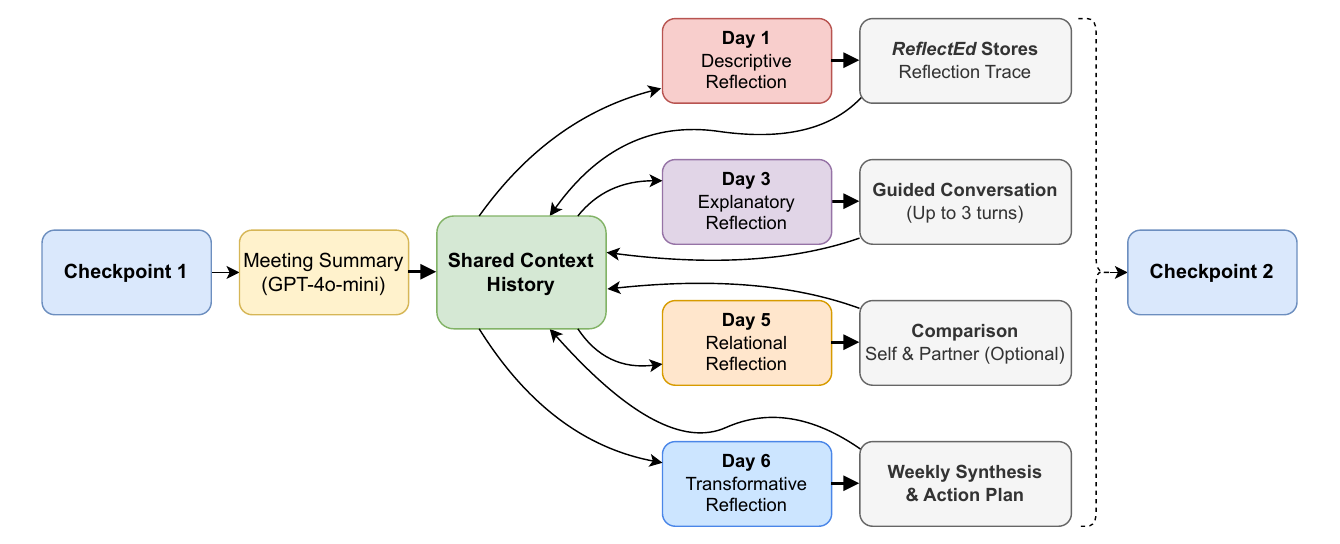}
    \caption{Overview of the multi-day reflection workflow. Each checkpoint meeting is summarized by GPT-4o-mini and incrementally enriches a shared context history across descriptive, explanatory, relational, and transformative reflection stages. \textbf{\textit{Note:}} Day~6 reflection is only applicable in the Deeper condition.}
    \label{fig:sys_navi}
\end{figure} 

\textit{\textbf{Participants and Procedure:}} We recruited 20 students (12 male, 8 female; ages 20–34, $M=26.40$, $SD=3.87$) from undergraduate and graduate programs across universities in the United States using convenience and snowball sampling. All participants had previously completed collaborative academic tasks at least twice in their academic or professional experience. Participants were expected to spend approximately 10 hours on the study and received \$80 in compensation. The study protocol was approved by the authors’ Institutional Review Board (IRB). Participants were randomly assigned to one of two conditions (five pairs per condition): Group~1 received the regular reflection workflow on alternate days between checkpoints (CPs), whereas Group~2 received the deeper reflection workflow on alternate days plus an additional Day~6 reflection activity. Each checkpoint corresponded to a scheduled meeting.

\begin{figure}[t]
\centering
\begin{subfigure}[t]{0.32\textwidth}
  \centering
  \includegraphics[width=\linewidth]{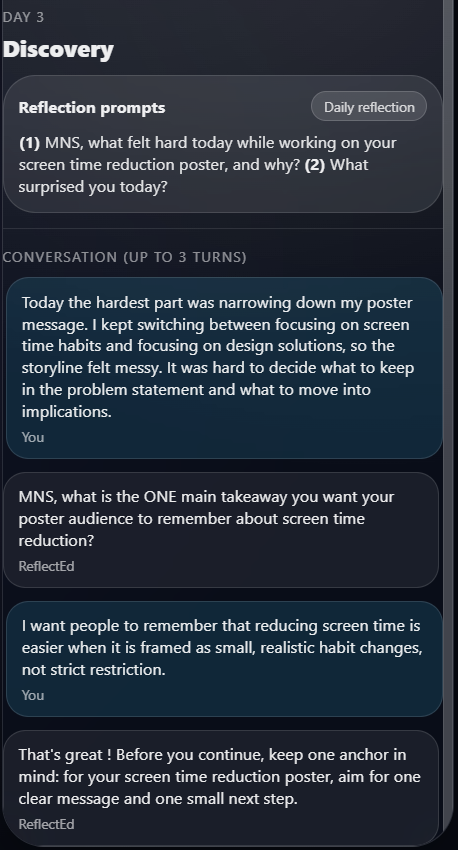}
  \caption{Day~3 (Regular and Deeper)}
\end{subfigure}
\hfill
\begin{subfigure}[t]{0.32\textwidth}
  \centering
  \includegraphics[width=\linewidth]{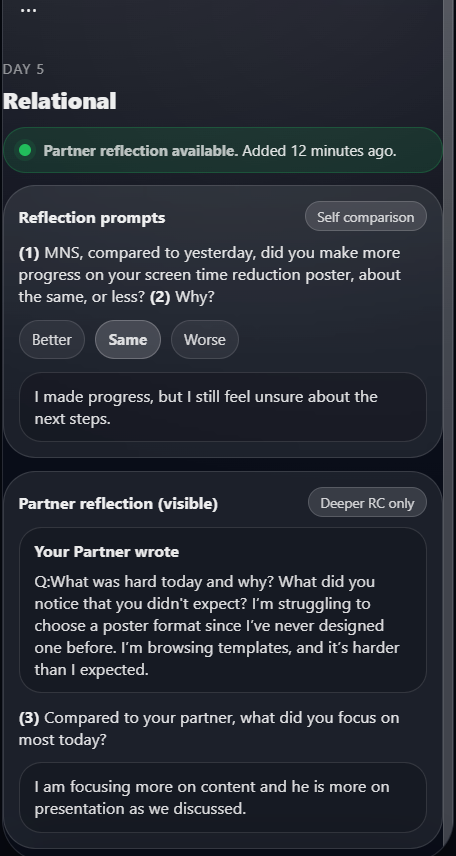}
  \caption{Day~5 (Regular and Deeper)}
\end{subfigure}
\hfill
\begin{subfigure}[t]{0.32\textwidth}
  \centering
  \includegraphics[width=\linewidth]{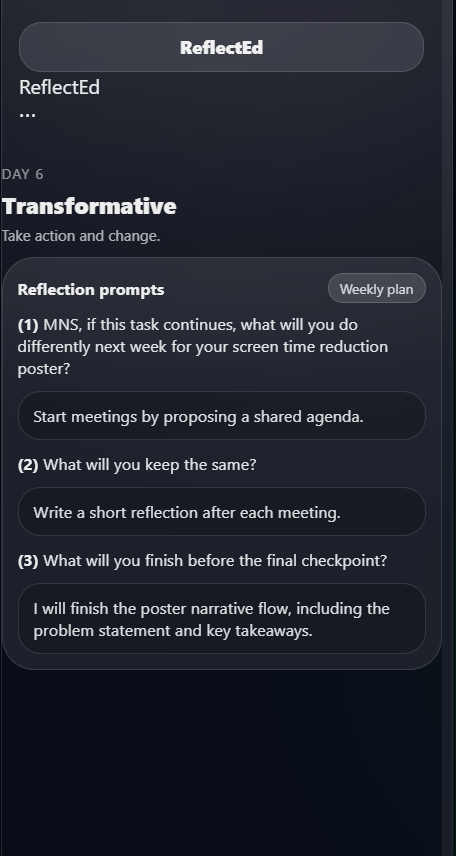}
  \caption{Day~6 (Deeper only)}
\end{subfigure}
\caption{Example snapshots from \mb~for Day~3, Day~5, and Day~6 reflection activities.}
\label{fig:reflectEd_snaps}
\end{figure}

\textit{\textbf{System Development:}}
\mb~is a web-based system for structured reflection across two checkpoints in a short collaborative workflow. It was built with Streamlit (front-end) and MongoDB (data storage). Meetings were conducted within \mb, and transcripts were automatically captured and processed via GPT-4o-mini API calls to generate meeting summaries and extract key discussion points for later reflection. We used one-shot and few-shot prompting to generate summaries, personalize reflective prompts, and guide interactions with \mb. Example snapshots of \mb~are shown in Fig. \ref{fig:reflectEd_snaps}.

\mb~runs an AI-assisted reflection workflow across seven days. Participants had individual account where they logged in daily after receiving notification about their reflection via email. At Checkpoint~1, they meet through \mb~and it generates an AI meeting summary followed by their meeting and maintains a compact shared context history storing users’ reflection traces and task progress. The workflow progresses from \textbf{Descriptive recap} (Day~1) to \textbf{Explanatory sensemaking} (Day~3), where \mb~conducts a bounded conversation of up to three turns (one targeted question with up to two follow-ups) based on the user’s latest entry. To keep the interaction short, non-interfering but guided, \mb~extracts lightweight cues (e.g., blocker, dependency), selects questions from a small task-grounded template set, and constrains follow-ups to clarification or next-action planning. On Day~5, \mb~supports \textbf{Relational reflection} through self-comparison across days and partner-comparison when available. On Day~6, it enables \textbf{Transformative reflection} by using accumulated context to generate a weekly synthesis (deeper condition) and actionable plan shown before Checkpoint~2 for meeting readiness. At Checkpoint 2, they again meet to finish up their task of designing the poster.

\textit{\textbf{Evaluation:}}
To evaluate \mb, we combined self-reports with trace- and artifact-based measures from reflections and checkpoint meetings. After the seven-day workflow, participants completed a post-study survey assessing (1) collaboration preparedness for CP2 (preparedness, clarity of progress, remaining tasks, and partner alignment), (2) reflection usefulness (sensemaking, identifying next actions, staying focused, and improving collaboration), and (3) system perception (ease of use, prompt relevance, structure, and workflow fit) using likert scale (Strongly Disagree-1 to Strongly Agree-5), and provided open-ended feedback analyzed thematically. We coded reflection traces using Kember et al.’s four-level framework \cite{kember2008four} (non-reflection, understanding, reflection, critical reflection), focusing comparisons on shared days (Day 1/3/5) since only the Deeper condition included Day 6. Because learning-to-action can appear more directly in planning than in critical reflection, we also coded actionability on a 0–2 scale (0 = not actionable, 2 = clearly actionable) and assessed follow-through by checking whether Day 6 plans were referenced in CP2. 

\section{Findings}
To analyze our findings, we primarily report descriptive quantitative patterns and use qualitative analysis to interpret open-ended responses. This mixed approach allows us to characterize how different reflection configurations relate to reflective engagement, feasibility, action planning, and preparedness, rather than to establish causal effects.

\subsection{Quantitative Analysis}

\begin{figure}[h!]
    \centering
    \includegraphics[width=1\linewidth]{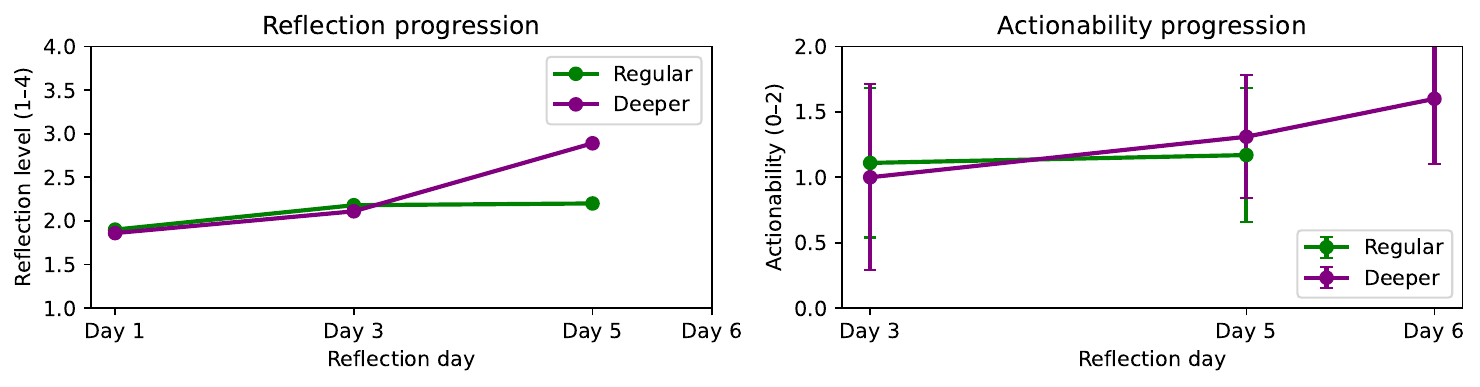}
    \caption{Reflection and actionability progression across days by condition.}
    \label{fig:result}
\end{figure}

\paragraph{\textbf{Reflection Progression:}}
We coded written reflections using a four-level scheme, achieving acceptable inter-rater reliability on a subset of the data (Cohen’s $k = 0.84$), after which disagreements were resolved through discussion. Early reflections showed similar levels across conditions, including Day~1 descriptive recaps (Regular: $M = 1.90$; Deeper: $M = 1.86$) and Day~3 explanatory reflections using the same prompt structure and bounded guidance (Regular: $M = 2.18$; Deeper: $M = 2.11$). By Day~5, reflections in the Deeper condition tended to reach higher coded levels (Deeper: $M = 2.89$; Regular: $M = 2.20$), coinciding with the introduction of partner-aware comparison. Across shared reflection days (Day~1, Day~3, Day~5), mean reflection levels were slightly higher in the deeper condition (Regular: $M = 2.09$, $SD = 0.41$; Deeper: $M = 2.25$, $SD = 0.44$), indicating a pattern of deeper reflective engagement under configurations that included additional comparison and synthesis features (Fig.~\ref{fig:result}).

\textit{\textbf{Feasibility and Cognitive Burden:}}
Across the seven-day workflow, the regular condition was associated with higher completion rates than the deeper condition. Participants completed a larger proportion of assigned reflections in the regular condition (Regular: $M = 0.97$, $SD = 0.10$; Deeper: $M = 0.80$, $SD = 0.22$; Table~\ref{tab:quant_summary}), suggesting that additional reflection depth and an extra reflection day were accompanied by lower adherence. The deeper condition was also associated with longer time-on-task per reflection (Regular: $M = 180$s, $SD = 32$s; Deeper: $M = 228$s, $SD = 46$s), reflecting an effort tradeoff that required more sustained attention during individual reflection activities.

\textit{\textbf{Actionability of Reflection Outcomes:}}
Actionability scores on Day~3 were comparable across conditions (Regular: $M = 1.11$, $SD = 0.57$; Deeper: $M = 1.01$, $SD = 0.71$), indicating similar levels of next-step planning following explanatory reflection and bounded \mb~guidance. On Day~5, reflections in the deeper condition showed slightly higher actionability scores (Regular: $M = 1.17$, $SD = 0.51$; Deeper: $M = 1.31$, $SD = 0.47$). For the Transformative reflection activity (deeper condition only), Day~6 reflections emphasized forward-looking planning, including intermediate deadlines, communication routines, and revised division of labor. These reflections exhibited moderate actionability (Day~6: $M = 1.46$, $SD = 0.50$ on a 0--2 scale). In addition, $82\%$ of participants in the deeper condition referenced at least one planned action during the second checkpoint meeting (CP2), suggesting that the Transformative reflection was frequently taken up as a resource for planning and meeting preparation (Fig.~\ref{fig:result}).

\begin{table}[t]
\centering
\caption{Feasibility, reflection progression (shared days: Day~1/3/5), actionability, and self-reported outcomes by condition. Values are mean (SD), except day-level reflection means which are reported as means.}
\label{tab:quant_summary}
\scriptsize
\setlength{\tabcolsep}{5pt}
\renewcommand{\arraystretch}{1.15}
\begin{tabular}{p{0.50\columnwidth} >{\centering\arraybackslash}p{0.20\columnwidth} >{\centering\arraybackslash}p{0.20\columnwidth}}
\hline
\textbf{Metric} & \textbf{Regular (n=10)} & \textbf{Deeper (n=10)} \\
\hline
Completion rate (assigned reflections) & 0.97 (0.10) & 0.80 (0.22) \\
Time-on-task per reflection (sec) & 180 (32) & 228 (46) \\
\hline
Reflection level (Day~1, mean) & 1.90 & 1.86 \\
Reflection level (Day~3, mean) & 2.18 & 2.11 \\
Reflection level (Day~5, mean) & 2.20 & 2.89 \\
Mean reflection level (shared days, 1--4) & 2.09 (0.41) & 2.25 (0.44) \\
\hline
Day~3 actionability (0--2) & 1.11 (0.57) & 1.01 (0.71) \\
Day~5 actionability (0--2) & 1.17 (0.51) & 1.31 (0.47) \\
Day~6 actionability (0--2) & -- & 1.46 (0.50) \\
Day~6 action carryover into CP2 (\%) & -- & 82\% \\
\hline
CP2 preparedness (self-report) & 3.84 (0.56) & 4.15 (0.51) \\
Reflection usefulness (self-report) & 4.06 (0.63) & 4.13 (0.58) \\
System perception (self-report) & 4.22 (0.43) & 4.08 (0.52) \\
\hline
\end{tabular}
\vspace{2pt}
\textit{\textbf{Note:} Reflection level coded using a four-level scheme (1--4). Actionability coded on a 0--2 scale. Self-reported outcomes used 5-point Likert scales (1--5). Day~6 metrics were collected in the deeper condition only.}
\end{table}

Participants’ post-study surveys indicated higher self-reported preparedness for the second checkpoint in the deeper condition (Regular: $M = 3.84$, $SD = 0.56$; Deeper: $M = 4.15$, $SD = 0.51$), while perceived reflection usefulness ratings were similar across conditions (Regular: $M = 4.06$, $SD = 0.63$; Deeper: $M = 4.13$, $SD = 0.58$). System perception ratings were positive and comparable overall (Regular: $M = 4.22$, $SD = 0.43$; Deeper: $M = 4.08$, $SD = 0.52$).

\subsection{Qualitative Analysis}
We analyzed open-ended post-study survey responses related to interface usability, reflection usefulness for planning and collaboration, partner visibility, second-checkpoint experiences, and improvement suggestions. Using grounded theory methods~\cite{thornberg2014grounded}, the first and second authors conducted open coding and iteratively refined themes through discussion. Initial agreement was high (Cohen’s Kappa = 0.86), and remaining differences were resolved collaboratively.

\textit{\textbf{Micro-checkpoints that sustain momentum:}}
Across both conditions, participants described reflection activities as lightweight pacing mechanisms that helped reduce drift between meetings and made progress more explicit. Participants valued brief pauses to review completed work and identify a next step. One participant described these check-ins as micro-deadlines: \textit{``When I reach the check-in, it feels like a small deadline\ldots{} I choose one next step and write it down\ldots{} and I begin with less hesitation.''} Participants also suggested pairing reflection prompts with simple task-support features in the same space, such as \textit{``a small checklist next to the prompt and a progress bar we both could see,''} to reduce coordination overhead and support shared awareness. Several participants noted that reflection reminders became particularly salient as CP2 approached, helping them re-engage with the task without feeling overwhelmed.

\textit{\textbf{Feasible reflection requires flexible depth:}}
While participants across conditions valued reflection, they also noted that deeper prompts (Day 6) could feel demanding on busy days. One participant in the deeper condition described this tradeoff: \textit{``On busy days, the deeper prompt (Day~6) felt a bit heavy, though it often led to useful reflection when I had the time.''} This feedback reflected a preference for mixed rhythms, in which lighter prompts supported continuity and deeper prompts supported sensemaking at key moments. As another participant summarized: \textit{``The factual question (Day~3) helped me remember my progress and keep procrastination in check\ldots{} That little conversation (Day~5) helped me see the difficulties\ldots{}''}. Together, these accounts highlight the importance of balancing reflection depth with feasibility in time-constrained workflows.

\textit{\textbf{Reflection-driven alignment supports readiness and transfer:}}
Participants in the deeper condition described partner-visible reflections as a coordination resource that helped clarify roles and supported accountability, particularly when preparing for the second checkpoint. Transformative reflection on Day~6 was frequently described as a way to consolidate weekly progress into concrete next steps, externalizing reasoning before re-entering collaborative work. This mattered during CP2, when teams needed to integrate work efficiently; some participants in the regular condition described spending time catching up on each other’s progress: \textit{``We spent the first half just figuring out what each of us had done\ldots{} we could have finished the poster faster.''} Beyond coordination, participants also noted that seeing differences in partners’ reflections surfaced alternative strategies: \textit{``It could also be surprising to see differences in how we approach the same task, which might give me new strategies.''} Participants in the deeper condition further reported that articulating planned changes before the next checkpoint helped them re-enter the task with clearer intent.

\section{Discussion}

\paragraph{\textbf{Design Tradeoffs in Reflection Depth:}}
A central implication of our findings is that reflection depth can be treated as a design parameter rather than a universal target. Both reflection configurations supported participants in maintaining momentum between checkpoints, but they differed in reflective depth, feasibility, and perceived preparedness. Early reflections were similar across conditions (Day~1 and Day~3), while differences in reflective patterns appeared later in the workflow. By Day~5, reflections in the deeper condition tended to reach higher coded levels and exhibited higher actionability, coinciding with the introduction of partner-aware comparison and later the Transformative reflection activity on Day~6. These patterns are consistent with accounts of transformative reflection, which emphasize changes in interpretation that inform subsequent action \cite{mezirow2000learning}. At the same time, increased depth was accompanied by reduced feasibility: participants in the deeper condition completed fewer reflections, often provided shorter responses, and spent more time per reflection, indicating higher burden during active work \cite{xu2025productive}.

Taken together, these results suggest that reflection-support systems may benefit from combining lightweight prompts that help maintain continuity with occasional deeper prompts placed at natural pauses, such as mid-cycle consolidation or pre-meeting planning. Such configurations can support reflection-on-action by helping learners connect experience to future decisions \cite{munby1989reflection}, without requiring sustained effortful elaboration throughout the workflow \cite{cho2022reflection}. In this framing, between-checkpoint reflection is not evaluated solely by the depth of written responses, but also by whether it helps learners articulate feasible next steps that can be carried forward into subsequent coordination moments \cite{gollwitzer1999implementation}. Consistent with this view, participants in the deeper condition reported higher perceived preparedness for the second checkpoint, while perceived usefulness and overall system perceptions remained similar across conditions.

Qualitative feedback further illustrated how reflection functioned as a form of ``micro-checkpoint'' that helped reduce drift between meetings, and highlighted the importance of flexible depth. Lighter prompts were described as helpful on busy days, while deeper prompts were perceived as most useful when participants had sufficient time and when partner-visible updates supported alignment by clarifying roles and next steps \cite{samrose2021meetingcoach,park2023retrospector}. At the same time, participants’ comments also underscored potential risks of visibility, such as impression management, reinforcing prior concerns that reflective technologies should preserve learner agency and avoid becoming surveillance-like \cite{baumer2015reflective,bentvelzen2022revisiting}. Design strategies such as selective sharing, where learners choose which commitments or blockers to surface, may help support coordination without making reflection feel performative \cite{roldan2021pedagogical}.

\paragraph{\textbf{Reflection as a Learning Scaffold:}}
Rather than demonstrating learning gains, our findings characterize how \mb~supports regulation processes that are commonly associated with learning in open-ended, project-based contexts. In academic project work, learners must interpret incomplete progress, diagnose stalls, and decide next steps with limited guidance, aligning with self-regulated learning accounts that emphasize monitoring, control, and adaptation \cite{barria2025using}. In this sense, between-checkpoint reflection is not only retrospective but also forward-looking, helping learners maintain awareness of progress, unresolved issues, and feasible actions as they move toward subsequent coordination points.

Participants described \mb~as providing lightweight conversational guidance that shaped how they re-engaged with their work, particularly as deadlines approached and effort intensified. Through different reflection stages, the system supported multiple regulation-oriented pathways: guiding learners when they surfaced challenges (feedback), enabling awareness of partner progress and differences (observation and comparison), and prompting revisitation of prior decisions (retrospection). These mechanisms were reflected in descriptive patterns of increasing actionability over time, as reflections more frequently articulated concrete next steps and forward-looking adjustments. Self-reports further suggested that deeper reflection prior to reconnecting with a partner was perceived as helpful for organizing work and preparing for the next checkpoint. At the same time, participants’ experiences indicated that the value of reflection depended on task demands, workload, and personal circumstances, suggesting that fixed reflection schedules may not be equally appropriate for all learners or contexts.

These findings also highlight attention as a prerequisite for productive reflection. Reflection appeared most supportive when learners could meaningfully attend to the task, consistent with prior work emphasizing the role of attention in learning processes \cite{dosher2010perceptual}. When learners were overloaded or engagement with the task was secondary, reflection was more likely to be experienced as burdensome rather than supportive. This points to the potential value of adaptive reflection designs that adjust depth and frequency based on task significance, deadline proximity, or learner readiness. Although our study examined a one-week workflow, similar principles may extend to longer-term projects by keeping reflection lightweight during routine execution while deepening near milestones \cite{pishtarireflectionapp}.

\paragraph{\textbf{Limitations and Future Work:}}
This study evaluated \mb~within a short, seven-day poster workflow with two checkpoints. While this design enabled focused examination of between-checkpoint reflection, it limits how directly the findings generalize to longer academic projects where meeting intervals vary, goals evolve, and progress is less structured. The study focused on pairs, and outcomes may differ in larger teams with more complex coordination dynamics or in instructor-mediated classroom settings. In addition, several outcomes relied on self-reported measures and coded reflection traces. These measures capture reflective engagement, feasibility, actionability, and perceived preparedness, but do not assess longer-term learning gains or transfer beyond the study period.

Future work should examine \mb~in longer deployments, explore adaptive reflection schedules that adjust prompt timing and depth in response to workload and milestones, and further investigate privacy-aware sharing mechanisms that support alignment without undermining autonomy or psychological safety. Together, these directions would help clarify how AI-assisted reflection can be integrated into authentic educational settings while respecting learners’ attention, agency, and evolving goals.

%
%
%
\bibliographystyle{splncs04}
\bibliography{reference}

\end{document}